\documentclass[graybox]{svmult}

\usepackage{helvet}         
\usepackage{courier}        
\usepackage{type1cm}        
\usepackage{makeidx}         
\usepackage{graphicx}        
\usepackage{multicol}        
\usepackage[bottom]{footmisc}
\usepackage{optidef}
\usepackage{graphicx}
\usepackage{amssymb}
\usepackage{color,soul}
\soulregister\cite7
\soulregister\ref7
\usepackage{lineno}
\usepackage[ruled,vlined,linesnumbered,titlenumbered]{algorithm2e}
\usepackage{times}
\usepackage{xurl}
\usepackage{hyperref}
\usepackage{optidef}
\usepackage{multirow}
\usepackage{booktabs}
\usepackage{enumitem}
{ \newcommand{\mynote}[2]{
     \fbox{\bfseries\sffamily\scriptsize#1}
       {\small$\blacktriangleright$\textsf{\textcolor{red}{{\em #2}\bf }}$\blacktriangleleft$}}}
       { \newcommand{\mynote}[2]{}}
\usepackage{tabularx,ragged2e,booktabs,caption}
\newcolumntype{C}[1]{>{\Centering}m{#1}}

\makeindex             

\begin{document}

\title*{QOPTLib: a Quantum Computing Oriented Benchmark for Combinatorial Optimization Problems}
\titlerunning{QOPTLib}
\author{Eneko Osaba and Esther Villar-Rodriguez}
\authorrunning{Osaba and Villar-Rodriguez}
\institute{Eneko Osaba (Corresponding Author), Esther Villar-Rodriguez \at TECNALIA, Basque Research and Technology Alliance (BRTA), 48160 Derio, Spain \email{eneko.osaba@tecnalia.com, esther.villar@tecnalia.com}}

\maketitle

\abstract{In this paper, we propose a quantum computing oriented benchmark for combinatorial optimization. This benchmark, coined as \texttt{QOPTLib}, is composed of 40 instances equally distributed over four well-known problems: Traveling Salesman Problem, Vehicle Routing Problem, one-dimensional Bin Packing Problem and the Maximum Cut Problem. The sizes of the instances in \texttt{QOPTLib} not only correspond to computationally addressable sizes, but also to the maximum length approachable with non-zero likelihood of getting a good result. In this regard, it is important to highlight that hybrid approaches are also taken into consideration. Thus, this benchmark constitutes the first effort to provide users a general-purpose dataset. Also in this paper, we introduce a first full solving of \texttt{QOPTLib} using two solvers based on quantum annealing. Our main intention with this is to establish a preliminary baseline, hoping to inspire other researchers to beat these outcomes with newly proposed quantum-based algorithms.}

\keywords{Quantum Computing, Optimization, Benchmarking, Quantum Annealing, DWAVE.}

\section{Introduction}
\label{sec:intro}

Quantum Computing is expected to provide researchers and practitioners with a novel and revolutionary paradigm for dealing with complex problems in a more efficient way \cite{nielsen2002quantum}. This kind of computation leverages quantum phenomena to find competent solutions via Quantum Processing Units (QPU). The potential of quantum computing has become apparent in several examples such as the Shor factorization algorithm \cite{shor1994algorithms} and the Grover quantum search algorithm \cite{grover1997quantum}. 

So far, there have been several research areas in which quantum computing has shown its potential, such as cryptography \cite{mavroeidis2018impact}, simulation \cite{tacchino2020quantum}, machine learning \cite{khan2020machine} or optimization \cite{li2020quantum}. This paper is focused on the last of this research streams. 

Quantum optimization has generated a profound impact in recent years. How to implement novel quantum solvers or how to introduce quantum methods in already existing classical pipelines or algorithms are currently widespread concerns in the community. In this regard, the fast advances in hardware technology \cite{ajagekar2019quantum} and the democratization in its access \cite{seskir2023democratization} have made research take off, especially in the optimization branch.

Regarding applications fields, transportation \cite{osaba2022systematic}, finance \cite{orus2019quantum}, energy \cite{ajagekar2019quantum} or medicine \cite{flother2023state} are some examples of how quantum optimization can contribute to the development of notable scientific advancements.

Even so, research cannot circumvent the state of the hardware. Current quantum computers suffer from certain limitations that directly affect their capability and performance. The current state of quantum computing is known as noisy intermediate-scale quantum (NISQ, \cite{Preskill2018}) era. Quantum devices available in this NISQ era are characterized by not being completely prepared to reliable deal with large problems.

This situation hinders the evaluation of quantum or hybrid methods being the researchers building their own benchmarks for each scientific proposal. In these very first steps in the quantum algorithmic designs a benchmark helps researchers to fairly assess their contributions. Iris dataset \cite{fisher1936use} or ImageNet \cite{deng2009imagenet} are excellent representatives of how a benchmark can be a cornerstone to gain momentum. 

Thus, this circumstance pushes researchers to generate their own testing benchmarks whenever they try to solve a particular problem using a quantum computer. This is the case even when dealing with well-known optimization problems, as can be seen in studies such as \cite{irie2019quantum,villar2022analyzing,azad2022solving,amaro2022case}, where authors built ad-hoc problem instances adapted to the limited capacity of quantum computers. This situation directly affects the replicability and comparison of technical approaches.

Taking this situation as main motivation, we present in this paper a quantum computing oriented benchmark for combinatorial optimization problems. This benchmark, coined as \texttt{QOPTLib}, is composed of 40 different instances equally distributed among four problems: the Traveling Salesman Problem (TSP, \cite{flood1956traveling}), the Vehicle Routing Problem (VRP, \cite{toth2002vehicle}), the one dimensional Bin Packing Problem (BPP, \cite{martello1990bin}), and the Maximum Cut Problem (MCP, \cite{bodlaender2000complexity}). Characteristics of each dataset are described in upcoming sections.

Also, we conduct a first experimentation in order to extract a preliminary baseline of results. For conducting these tests, two commercial solvers provided by DWAVE have been employed: a pure QPU based \texttt{Advantage\_system6.1}, and the quantum-classical \texttt{LeapHybridBQMSampler}. For each instance of \texttt{QOPTLib}, the best solution found by both solvers after 10 independent runs is given.

Our main objective with this study is to propose a benchmark of well-known combinatorial optimization problems as the key element of a testbed in quantum optimization.


The rest of this paper is organized as follows: in Section \ref{sec:problems}, we introduce the problems that have been considered in \texttt{QOPTLib}, highlighting briefly the related work done in each of them in the field of quantum computing. After that, Section \ref{sec:benchmark} deeply describes \texttt{QOPTLib}. We conduct in Section \ref{sec:exp} a preliminary experimentation with a pure QPU and a quantum-classical solver. This paper finishes in Section \ref{sec:conc} with conclusions and further work.

\section{Description of the problems}
\label{sec:problems}

As mentioned in the introduction, \texttt{QOPTLib} contemplates instances regarding four combinatorial optimization problems. This section is devoted to briefly describe each of these problems.

\subsection{Traveling Salesman Problem}
\label{sec:TSP}

The TSP is one of the most widely studied problems in operations research and computer science. Despite being a classical optimization algorithm, the TSP is the focus of many research works even today \cite{cheikhrouhou2021comprehensive,osaba2020traveling}, since transportation problems are still in the hype, and because of the analogy between many real world problems and the TSP formulation. As a result of this interest, the TSP is frequently used as benchmarking problem for testing the quality of newly proposed techniques and solvers \cite{liu2023daaco,bogyrbayeva2023deep}; and even for tackling real-world oriented transportation problems \cite{berczi2023approximations,kloster2023multiple}.

The TSP is defined as a complete graph $G= (V,A)$, where $V= \{v_1,v_2,\dots,v_n\}$ is the group of nodes and $A= \{(v_i,v_j): v_i,v_j  \in V,i\neq j \}$ is the set of links among these nodes. Additionally, traveling from one node to another has an associated cost $c_{ij}$. This cost is the same regardless of the direction, i.e. $c_{ij}$ = $c_{ji}$. Thus, the objective is to find a path which visits each and every node once while minimizing the total cost of the route. Also, the trip must start and finish at the same point. The TSP can be mathematically formulated as follows:

\begin{mini!}|l|
	{\mathbf{X}}{f(\mathbf{X}) = \sum_{i=1}^n \sum\limits_{\substack{j=1\\i\neq j}}^n c_{ij} x_{ij}\label{TSPeq1}}{}{}
	\addConstraint{\sum\nolimits_{\substack{j=1\\i\neq j}}^n x_{ij}}{=1,}{\quad\forall{j} \in\{1,\ldots,n\}\label{TSPeq3}}
	\addConstraint{\sum\nolimits_{\substack{i=1\\i\neq j}}^n x_{ij}}{=1,}{\quad\forall{i} \in\{1,\ldots,n\}\label{TSPeq4}}
	\addConstraint{\sum\nolimits_{\substack{i\in{\mathcal{S}}\\j\in{\mathcal{S}}\\i \neq j}} x_{ij}}{\geq 1,}{\quad\forall{\mathcal{S} \subset \mathcal{V},}\label{TSPeq5}}
\end{mini!}
where $x_{ij}\in\{0,1\}$ takes value $1$ if edge $(i,j)$ is used in the solution. Additionally, the objective function is depicted in Formula \eqref{TSPeq1} as the sum of all costs associated to the edges that compose the route. Furthermore, Restrictions \eqref{TSPeq3} and \eqref{TSPeq4} represent that each node must be visited once and only once. Finally, \eqref{TSPeq5} assures the non-existence of sub-tours, requiring that any subset of nodes $S$ has to be abandoned at least once.

Being such an interesting problem, the TSP has been one of the first combinatorial optimization problems to be solved by a QPU \cite{osaba2022systematic}, and many works have followed the lead exploring new solving schemes \cite{osaba2021hybrid,srinivasan2018efficient}; improving the formulation or proposing novel variants \cite{salehi2022unconstrained,osaba2021focusing}; or making use of the TSP to solve similar, from the mathematical perspective, real-world use cases \cite{mehta2019quantum,clark2019towards}.


Despite this interest, few advances have been made to create a TSP benchmarking, complicating the replicability and fair evaluation of the research published. 


\subsection{Vehicle Routing Problem}
\label{sec:VRP}

Like the TSP, the VRP is one of the most renowned and intensively studied problems in artificial intelligence field. In a nutshell, the VRP is an extension of the TSP, in which the objective is to find the optimal route planning for a fleet of vehicles, given the demands of a set of $n$ client. Thus, the problem can be defined in a similar way to the TSP, as a complete $G= (V, A)$ where $V$ represent the set of clients and $A$ the set of connections among them, having each edge $(v_i,v_j)$ an associated cost $c_{ij}$ (in this case, also $c_{ij}$ = $c_{ji}$). Additionally, in the case of the VRP, the node $v_0$ represents a depot, in which all vehicles should start and end their route. The rest of nodes represent the clients to serve according to their demands $q_i$, which are grouped in $Q = \{q_1, q_2, \dots , q_n \}$.

Lastly, each VRP instance counts with a fleet of vehicles $K$. Each available vehicle has a limited $C$ capacity. Depending on the VRP variant, the size of $K$ could be limited or unlimited, and the use of a certain vehicle could imply an additional cost to the route. In any case, in the majority of the VRP formulations, the fleet is composed of an unlimited number of free-using identical vehicles.

With all this, the main objective of the canonical VRP is to find a number of routes minimizing the total cost, considering that \textit{i)} each trip must start and end at $v_0$, \textit{ii)} each client must be visited once and \textit{iii)} the accumulated demand satisfied by each route does not exceed the vehicle capacity \cite{CVRPform}. Mathematically, the VRP can be formulated as follows \cite{borcinova2017two}:

\begin{mini!}|l|
	{\mathbf{X}}{f(\mathbf{X}) = \sum_{r=1}^p \sum_{i=1}^n \sum\limits_{\substack{j=1\\i\neq j}}^n c_{ij} x_{rij}\label{VRPeq1}}{}{}
	\addConstraint{\sum_{\substack{r=1}}^n \sum_{\substack{j=1, i\neq j}}^n x_{rij}}{=1,}{\quad\forall{j} \in\{1,\ldots,n\}\label{VRPeq2}}
    \addConstraint{\sum_{\substack{j=1}}^n x_{r1j}}{=1,}{\quad\forall{r} \in\{1,\ldots,p\}\label{VRPeq3}}
    \addConstraint{\sum_{\substack{i=1 \\ i\neq j}}^n x_{rij}=\sum_{\substack{i=1}}^n x_{rji},}{\quad\forall{j} \in\{1,\ldots,n\},{r} \in\{1,\ldots,p\}\label{VRPeq4}}
    \addConstraint{\sum_{\substack{i=1}}^n \sum_{\substack{j=1, i\neq j}}^n q_j x_{rij}}{< C,}{\quad\forall{r} \in\{1,\ldots,p\}\label{VRPeq5}}
    \addConstraint{\sum_{\substack{r=1}}^p\sum_{\substack{i \in \mathcal{S}}} \sum_{\substack{j \in \mathcal{S}, i\neq j}} x_{rij}}{\leq |S|-1,}{\quad\forall{S} \subseteq\{1,\ldots,n\}\label{VRPeq6}}
\end{mini!}
where the binary variable $x_{rij}$ is 1 if edge ($i,j$) is part of the route $r$, and $p$ is the number of paths that compose the whole solution. Formula \ref{VRPeq1} represents the objective function. Restriction \ref{VRPeq2} guarantees that each customer is visited once. Constraints \ref{VRPeq3} and \ref{VRPeq4} assure that each vehicle leaves the deport, and that the number of vehicles arriving to each customer is equal to the amount of vehicles leaving it. Formula \ref{VRPeq2} regards the capacity restriction, while Constraint \ref{VRPeq6} ensures the non-existence of sub-tours.

The research conducted on the VRP in last decades has been extraordinarily prolific \cite{konstantakopoulos2022vehicle,osaba2020vehicle}, mainly because of the flexibility of the problem to be adapted to a wide variety of real-world settings. As a result of this abundant activity, a plethora of VRP variants have been proposed in the literature, considering different realistic features such as time windows \cite{braysy2005vehicle}, heterogeneous vehicles \cite{kocc2016thirty}, multiple depots \cite{min1992multiple}, or pickup and deliveries \cite{parragh2008survey}. In addition to that, a specific group of variants coined as Rich VRP \cite{caceres2014rich,lahyani2015rich} has emerged, which refers to these VRP variant contemplating an elevated number of constraints \cite{yang2020cooperative}.

One immediate consequence of this increasing community is the vast amount of ad-hoc benchmarks. Some of these VRP datasets have become standards for the community, and they are often used as baseline for testing the quality of newly proposed methods. Some examples are the benchmark of Solomon \cite{solomon1987algorithms}, Cordeau \cite{cordeau2002vehicle}, Christofides and Eilon \cite{christofides1969algorithm}, or Fisher \cite{fisher1994optimal}.

Turning our attention to quantum computing, some interesting research has also been conducted in this field, although practical studies are still scarce \cite{osaba2022systematic}. In \cite{feld2019hybrid}, for example, authors 
present a two-step method consisting of a clustering algorithm to estimate the number of vehicles to later solve each route as a TSP problem by the DWAVE. Further hybrid solvers can be found in \cite{borowski2020new} or \cite{mohanty2022analysis}. Also, pure quantum solvers have also been proposed in the last years, as can be seen in \cite{irie2019quantum}, \cite{azad2020solving} or \cite{harwood2021formulating}. In any case, it should be considered that, even for hybrid schemes such as the one proposed in \cite{feld2019hybrid}, most works do the experimentation on small problems, and those targeting larger instances are just centered on formulating different kind of VRP variants \cite{harikrishnakumar2020quantum,syrichas2017large} in a theoretical framework.


\subsection{Bin Packing Problem}
\label{sec:BPP}

The packing of items (or packages) into a minimum number of containers (or bins) is a common task in logistics and transportation systems. In the field of operation research, this problem, known as Bin Packing Problem (BPP), has been in the spotlight for the last years. Depending on the characteristics of both items to store and bins, different variants of this problem can be formulated. Arguably, the simplest BPP formulation is the so-called one-dimensional BPP (1dBPP, \cite{munien2021metaheuristic}), which consists of a set of items $I \in \{i_1, i_2,\dots,i_n\}$ only defined by their weights $w_i$, and an unlimited number of bins with a maximum capacity $C$. 

On top of this simple formulation, more complex problems have been introduced in the literature \cite{delorme2016bin}: such as the two-dimensional BPP (2dBPP, \cite{lodi2014two}) to define height and width of items and bins, and the three-dimensional BPP (3dBPP, \cite{martello2000three}), in which the depth is also considered. Besides, all these BPP formulations can be extended to meet new requirements such as fragility \cite{el2020heuristic}, time windows \cite{liu2021algorithms} or compatibilities \cite{santos2019variable}.

Anyway, even though BPP is still the kingpin of many industrial processes, it was not until 2022 that the first research purely focused on this kind of problems was conducted by the quantum community \cite{de2022hybrid,garcia2022comparative}. 
Besides those two papers, in \cite{bozhedarov2023quantum} we can find a study tackling with the problem of filling spent nuclear fuel in deep-repository canisters. Authors formulate this problem as a 1dBPP, solving it using the D-Wave quantum annealer. More recently, authors in \cite{romero2023hybrid} explore the first solving algorithm for a 3dBPP through a hybrid quantum solver. 

Finally, similarly to what occurs with the aforementioned problems, no standard benchmarking for quantum solvers exists, in spite of the fact that there are public datasets such as the BPPLib \cite{delorme2018bpplib}. This situation contributes to the value of \texttt{QOPTLib}, which includes 10 instances of the 1dBPP.


\subsection{Maximum Cut Problem}
\label{sec:MCP}

The Maximum Cut Problem, also known as Max-Cut, is a combinatorial optimization problem which principal objective is to divide the vertices of a graph into two disjoint subsets, in such a way that the number (or the sum of weights in a weighted graph) of crossing edges between these two subgroups is maximum. Proven to be an NP-Hard \cite{karp1972reducibility} problem, the MCP has been applied up to now to diverse fields such as network science \cite{ghatee2013hopfield} or clustering \cite{ding2001min}.

The MCP can be formally defined as an undirected graph $G= (V, A)$, where $V$ is the set of nodes and $A$ the set of edges. Also, the weight of the edge linking $v_i$ and $v_j$ is defined as $w_{i,j}$. Thus, a cut ($S$,$S'$) is a partition of $V$ into two subgraphs $S$, $S' = V/S$. The value of this cut is calculated as the sum of the weights $w_{ij}$ of the edges that connect the two disjoint subsets $S$ and $S'$. As explained before, the goal of the MCP is to maximize the value of a cut.

The MCP has been the focus of a myriad number of works along the last decades \cite{dunning2018works}, giving rise to a wide variety of datasets. In this context, the Stanford university published \texttt{GSet} as a standard format for creating problem instances\footnote{https://web.stanford.edu/~yyye/yyye/Gset/}. In addition, many quantum algorithms have been applied to the resolution of MCPs, like QAOA \cite{villalba2021improvement,crooks2018performance,guerreschi2019qaoa} and quantum annealers \cite{hamerly2019experimental}. Due to the importance of MCP in the bibliography, it is a good practice to include it in our benchmark.

\section{Introducing the generated \texttt{QOPTLib} benchmarks}
\label{sec:benchmark}

With all this, we introduce \texttt{QOPTLib} in this paper, which is a benchmark comprised of 40 different instances, evenly distributed among the problems described in Section \ref{sec:problems}: TSP, VRP, 1dBPP and MCP. \texttt{QOPTLib} is openly available under demand, or online at \cite{QOPTLib}. 

The problem sizes have been selected to comprehend both small toy samples and more complex, yet approachable, instances to be solved by a QPU. Furthermore, the biggest instances have been empirically designed according to the capacity of the hybrid methods provided by DWAVE and the practical problem size with a non-zero probability of achieving good results. In this regard, DWAVE has been considered for this study since it is the platform that accepts the largest problem sizes. 

Being more specific, it is known that current quantum annealers, such as DWAVE, can deal with larger problems in comparison with quantum gate based methods, such as QAOA \cite{atchade2021qrobot,mugel2022dynamic}. Similarly, in the context of quantum annealers, commercial hybrid approaches can deal with even bigger problem with respect to purely quantum based alternatives \cite{cohen2020portfolio}.

Thus, our goal is to create a benchmark with a number of instances able to evaluate any quantum solver in any QPU. In addition to this, the instances built should suppose a challenge for both the most restrictive and the most advanced algorithms. Now, we proceed to describe the four different datasets included in our benchmark.

\begin{figure}[t]
	\centering
	\includegraphics[width=0.55\linewidth]{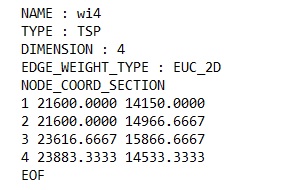}
	\caption{Structure of \texttt{wi4} TSP instance.}
	\label{fig:wi4}
\end{figure}

\textbf{Benchmarking instances for the TSP}: the dataset generated for the TSP is composed of 10 different instance with sizes ranging 4 nodes to 25. For the generation of this dataset, two well-known \texttt{TSPLib} instances have been selected as base: \texttt{wi29} and \texttt{dj38}. These original instances have been then reduced to produce 10 different use cases: \texttt{wi4}, \texttt{wi5}, \texttt{wi6}, \texttt{wi7}, \texttt{wi25}, \texttt{dj8}, \texttt{dj9}, \texttt{dj10}, \texttt{dj15} and \texttt{dj22}. To keep track of the original instance in \texttt{TSPLib}, each instance has been named as \texttt{djX} and \texttt{wiX}, where $X$ is the size of the problem. Regarding the format, it is \texttt{TSPLib} standard compliant, being readable by existent libraries, such as the \texttt{Qiskit TSP}\footnote{https://qiskit.org/documentation/optimization/stubs/qiskit\_optimization.applications.Tsp.html}. Figure \ref{fig:wi4} illustrates the structure of \texttt{wi4} instance.

\begin{figure}[h]
	\centering
	\includegraphics[width=0.5\linewidth]{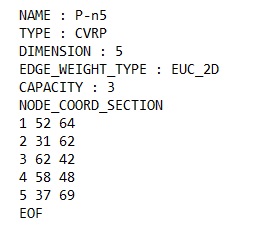}
	\caption{Structure of \texttt{P-n5} VRP instance.}
	\label{fig:P-n5}
\end{figure}

\textbf{Benchmarking instances for the VRP}: 10 newly created instances of the VRP have been included in \texttt{QOPTLib}, composed of 4 to 8 nodes. All these cases are reductions of the well-known \texttt{P-n16-k8} and \texttt{P-n23-k8} instances from the Augerat CVRP benchmark \cite{augerat1998separating}. Furthermore, in order to relax the complexity of the problem, the demands $q$ of all clients have been set to 1, and the vehicle capacity $C$ has been choosen to ensure that the optimum solution contains only two routes. Each instance has been named as \texttt{P-nX\_Y}, where $X$ is the number of nodes and $Y$ is the suffix to distinguish the set of instances with same $X$. Additionally, the VRP dataset built is \texttt{CVRPLib}\footnote{http://vrp.galgos.inf.puc-rio.br/index.php/en/} compliant. We depict in Figure \ref{fig:P-n5} an example of instance, specifically focused on \texttt{P-n5\_1}. 

\textbf{Benchmarking instances for the 1dBPP}: in \texttt{QOPTLib} 10 cases for the 1dBPP have been incorporated. In this case, these 10 instances have been crafted by the authors of this paper, following the format described in already published studies such as \cite{de2022hybrid} and \cite{garcia2022comparative}. They have been randomly configured while keeping set boundaries: from 3 to 14 for number of packages; \{10,12,15\} for bin capacities and 2 or 5 as for the values of package weights. Additionally, the number of bins available on each instance is the same as the number of packages that compose it.  

\begin{figure}[h]
\centering
\includegraphics[width=0.15\linewidth]{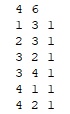}
\caption{Structure of an MCP instance composed of 4 nodes.}
\label{fig:MaxCut}
\end{figure}

\textbf{Benchmarking instances for the MCP}: the dataset included in \texttt{QOPTLib} for the MCP is composed of 10 instances with sizes ranging from 10 to 300 nodes. Each case has been coined as \texttt{MaxCut\_X}, being $X$ the cardinality of graph $G$. All these instances have been randomly generated through a Python script which has been implemented for this specific research. This instance generation script is also openly available in the Mendeley Data repository \cite{QOPTLib}. It is also interesting to mention that all use cases have the standard format \texttt{GSet}, which eases their automatic reading by already implemented libraries\footnote{https://qiskit.org/documentation/optimization/stubs/qiskit\_optimization.applications.Maxcut.html}. Figure \ref{fig:MaxCut} describes the structure of an instance with thess parameters: cardinality of $G$ and total sum of weights (first line), edge $\{i,j,w\}$ (the rest of the lines), where $i$ is the origin node, $j$ the destination node, and $w$ the weight of this link.

\section{Preliminary experimentation}
\label{sec:exp}

Despite not being the main objective of this paper, we present in this section a first complete set of results for \texttt{QOPTLib}. For this purpose, we resort to the DWAVE system, embracing two different solving approaches. On the one hand, we use the \texttt{Advantage\_system6.1} as a quantum pure alternative. This computer has 5616 working qubits, which are distributed using a Pegasus topology. On the other hand, we employ the \texttt{LeapHybridBQMSampler}, which is a hybrid solver of DWAVE to deal with general, often larger, binary quadratic problems.

\begin{table}[t!]
	\centering
	\resizebox{0.9\columnwidth}{!}{
		\begin{tabular}{C{0.5in} C{0.5in} C{0.5in} | C{0.75in} C{0.5in} C{0.75in} }
			\toprule[1.5pt]

                \multicolumn{3}{c | }{\bf Traveling Salesman Problem} & \multicolumn{3}{c}{ \bf Bin Packing Problem}\\\midrule
                \multirow{2}{*}{\bf Instance} & \multicolumn{2}{c|}{\bf Results} & \multirow{2}{*}{\bf Instance} & \multicolumn{2}{c}{\bf Results}\\
                
                 & QPU & Hybrid & & QPU & Hybrid \\
                
                \midrule
                \texttt{wi4}  & 6700  & 6700 & \texttt{BPP\_3} & 2 & 2 \\
                \texttt{wi5}  & 6786  & 6786 & \texttt{BPP\_4} & 2 & 2 \\
                \texttt{wi6}  & 9815  & 9815 & \texttt{BPP\_5} & 3 & 2 \\
                \texttt{wi7}  & 7267  & 7245 & \texttt{BPP\_6} & \texttt{--} & 3\\
                \texttt{dj8}  & 3269  & 2794 & \texttt{BPP\_7} & \texttt{--} & 4\\
                \texttt{dj9}  & \texttt{--} & 2438 & \texttt{BPP\_8} & \texttt{--} & 4\\
                \texttt{dj10}  & \texttt{--} & 3155 & \texttt{BPP\_9} & \texttt{--} & 4\\
                \texttt{dj15}  & \texttt{--} & 5268 & \texttt{BPP\_10} & \texttt{--} & 6\\
                \texttt{dj22}  & \texttt{--} & 13005 & \texttt{BPP\_12} & \texttt{--} & 7\\
                \texttt{wi25}  & \texttt{--} & 83132 & \texttt{BPP\_14} & \texttt{--} & 7\\
                
                \midrule
                \multicolumn{3}{c | }{\bf Vehicle Routing Problem} & \multicolumn{3}{c}{\bf Maximum Cut Problem}\\\midrule
                \multirow{2}{*}{\bf Instance} & \multicolumn{2}{c|}{\bf Results} & \multirow{2}{*}{\bf Instance} & \multicolumn{2}{c}{\bf Results}\\
                 & QPU & Hybrid & & QPU & Hybrid \\
                \midrule
                \texttt{P-n4\_1} & 97  & 97 & \texttt{MaxCut\_10} & 25 & 25 \\
                \texttt{P-n4\_2} & 121  & 121 & \texttt{MaxCut\_20} & 97 & 97 \\
                \texttt{P-n5\_1} & 120 & 94 & \texttt{MaxCut\_40} & 355 & 355 \\
                \texttt{P-n5\_2} & 315  & 295 & \texttt{MaxCut\_50} & 560 & 602\\
                \texttt{P-n6\_1} & \texttt{--} & 118 & \texttt{MaxCut\_60} & 756 & 852\\
                \texttt{P-n6\_2} & \texttt{--} & 122 & \texttt{MaxCut\_100} & \texttt{--} & 2224\\
                \texttt{P-n7\_1} & \texttt{--} & 119 & \texttt{MaxCut\_150} & \texttt{--} & 4899\\
                \texttt{P-n7\_2} & \texttt{--} & 164 & \texttt{MaxCut\_200} & \texttt{--} & 8717\\
                \texttt{P-n8\_1} & \texttt{--} & 153 & \texttt{MaxCut\_250} & \texttt{--} & 13460\\
                \texttt{P-n8\_2} & \texttt{--} & 269 & \texttt{MaxCut\_300} & \texttt{--} & 19267\\
            \bottomrule[1.25pt]
		\end{tabular}
	}
	\caption{Best results obtained by \texttt{Advantage\_system6.1} (QPU) and \texttt{LeapHybridBQMSampler} (Hybrid) for the whole \texttt{QOptLib} benchmark. \texttt{--} represents that the instance is unmanageable for the solver.}
	\label{tab:results}
\end{table}

Having said this, Table \ref{tab:results} shows the results obtained by both approaches. As mentioned, hybrid solver of leap allows addressing larger problems. Consequently, this hybrid scheme can deal with the whole \texttt{QOPTLib}, whereas the pure quantum \texttt{Advantage\_system6.1} can only cope with a smaller proportion of the entire benchmark due to its limited capacity. For each problem instance, 10 independent runs have been executed. For each case, we depict the best results obtained as a baseline for future studies. Furthermore, Symbol \texttt{--} represent 'no outcome' when the instance is unprocessable for the pure quantum approach.

\section{Conclusions and further work}
\label{sec:conc}

In the field of quantum optimization, the evaluation of novel resolution methods often faces the problem of the non-existence of benchmarks as standards for fair comparisons.
The community working on TSP and BPP, for example, relies on well-know libraries, such as \texttt{TSPLib} or the \texttt{BPPLib}. However, these benchmarks are not suitable for current quantum devices.
This situation pushes researchers to create their own instances adapted to the quantum computers they are using. This has a significant impact on the replicability of the studies and the generation of common knowledge. 

With this motivation in mind, we have introduced \texttt{QOPTLib} in this paper, which is a quantum computing oriented benchmark for combinatorial optimization. \texttt{QOPTLib} includes 40 different instances, equally distributed among four well-known problems: Traveling Salesman Problem, Vehicle Routing Problem, one-dimensional Bin Packing Problem and the Maximum Cut Problem. In addition to that, we have also presented a first complete solving of \texttt{QOPTLib} using two commercial solvers provided by DWAVE: the pure quantum \texttt{Advantage\_system6.1}, and the \texttt{LeapHybridBQMSampler}. The principal objective is not find the optimality of \texttt{QOPTLib}, but to provide with a baseline results, hoping that future works beat these outcomes with newly proposed algorithms.

Our team has now the sights set on two main research lines. In the short-term, we plan to increase the scope of \texttt{QOPTLib}, aggregating further instances to accommodate additional combinatorial optimization problems, such as the Job-Shop Scheduling Problem \cite{applegate1991computational} or other problems related with economics \cite{egger2020quantum} or energy \cite{szedlak2022extended} fields. In the medium-term, we will also propose new quantum procedures tested on \texttt{OPtLib}, with the main goal of assessing newly proposed quantum computing based solvers.

\begin{acknowledgement}
This work was supported by the Basque Government through ELKARTEK program (BRTA-QUANTUM project, KK-2022/00041), and through HAZITEK program (Q4\_Real project, ZE-2022/00033). This work was also supported by the Spanish CDTI through Plan complementario comunicación cuántica (EXP. 2022/01341)(A/20220551).
\end{acknowledgement}

\bibliographystyle{splncs}
\bibliography{biblio}

\end{document}